\documentclass[12pt]{article}
\usepackage{graphicx}
\begin{document}
{\bf Physics of randomness and regularities for cities, languages,}

{\bf and their lifetimes and family trees}

\medskip

\centerline{\c{C}. Tuncay$^1$}

\medskip
\centerline{$^1$ Department of Physics, Middle East Technical
University}

\centerline{06531 Ankara, Turkey, e-mail: caglart@metu.edu.tr}

\medskip
Abstract: Time evolution of the cities and of the languages is
considered in terms of multiplicative noise \cite{[1]} and
fragmentation  \cite{[2]} processes; where power law (Pareto-Zipf
law) \cite{[3]} and slightly asymmetric log-normal (Gauss)
\cite{[4]} distribution result for the size distribution of the
cities and for that of the languages, respectively. The cities and
the languages are treated differently (and as connected; for
example, the languages split in terms of splitting the cities, etc.)
and thus two distributions are obtained in the same computation at
the same time. Evolutions of lifetimes and families for the cities
and the languages are also studied. We suggest that the regularities
may be evolving out of randomness, in terms of the relevant
processes.

\section{Introduction}

Our essential aim is to show that time evolution of the competition
between cities or languages might have been governed by two opposite
processes; namely, random multiplicative noise for growth in size
and fragmentation for spread in number and extinction.   The
following section is the model, and the next one is the applications
and results. Last section is devoted for discussion and conclusion.

\bigskip
\section{Model}

This section is the description of the multiplicative noise
\cite{[1]} and the fragmentation \cite{[2]} processes for the
evolution of the competition between the cities or the languages,
where we introduce the meaning of the relevant concepts and the
parameters (with the symbols in capital letters for the cities and
those in lower case for the languages).

\subsection{Initial world}

Initially we have $M(0)$ ancestors for the cities ($I$) and $m(0)$
ancestors for the languages ($i$), where each city has a random size
$P_I(0)$ and she speaks one of the initial languages, which is
selected randomly. So, $P_I(0)$ is the population of each ancestor
city, and $p_i(0)$ is the number of people speaking each ancestor
language.

\subsection{Processes governing the evolution of the cities and this of
the languages}

Cities grow in time $t$, with a random rate $R_I \leq R$, where $R$
is universal within a \textit {random multiplicative noise process},

$$P_I(t) = (1 + R_I)P_I(t - 1)   .                   \eqno(1)$$
It is obvious that, as the initial cities grow in population the
initial languages grow in size $p_i(t)$; however, the cities {\it
fragment} in the meantime: Instead, if a random number (defined
differently at each time step $t$) for a city is smaller than the
splitting probability $H$, then the city splits after growing, with
the splitting ratio (fragmentation, mutation factor) $S$, where the
fragmentation (splitting, mutation) has the following meaning: If
the current number of habitants of a city $I$ is $P_I$($t$),
$SP_I$($t$) many members form another population and $(1-S)P_I(t)$
many survive within the same city; here, the results do not change
if $1 - S$ is substituted for $S$, i.e., if the mutated and
surviving members are interchanged. The number of the cities
$M$($t$) increases by one if one city splits; if any two of them
split at $t$, then $M$($t$) increases by two, etc. Furthermore, we
assume that if for $H  \ll 1 $ a random number is larger than some
$G \approx 1$, the city becomes extinct (random elimination,
punctuation).  A newly generated city speaks with probability $h_f$
a new language, with probability $h_s$ the current language of a
randomly selected city, and with the remaining probability 1 - $h_f$
- $h_s$ the old language of the mother city. This finishes the
definition of the model.

If a new city forms a new language ($h_f$) then it means that the
language of the home city is fragmented; where, the splitting ratio
$S$ is the ratio of the population of this new city to the total
population of the cities which speak the old language. It is obvious
that $h_f$ is small ($h_f \ll 1$); yet, many new languages may
emerge at each time step, since many new cities emerge in the mean
time, and $h_f$ becomes important. On the other hand, a new (and an
old) city may change her language and select one of the current
languages as the new one (with $h_s \approx 0$, for all), where
colonization may take place or teachers may teach the new language
\cite{[4]}, etc. In this case, the size of the old (new) language
decreases (increases) by the population of the new city. It is
obvious that the language which is spoken by many cities has a
higher chance for being selected by a new city; and so, big
languages are favored in case of selecting a new language. And, many
new cities continue speaking the language of the home city (country,
state) at the present era.

Please note that the fragmentation causes new cities or languages to
emerge (birth), and at the same time it drives them to extinction in
terms of splitting, and any city or a language with a member less
than unity is considered as extinct. The number of the cities
increases, decreases, or fluctuates about $M(0)$ for relatively big
numbers for $H$ (high fragmentation) and $G$ (low elimination), for
small numbers for $H$ (low fragmentation) and $G$ (high
elimination), and for $H$ + $G$ = 1 (equal fragmentation and
elimination), respectively; out of which we regard only the first
case, where we have (for $1< H + G$) an increase in the number of
cities, and we disregard the others. For it is known (guessed) that
the number of the cities did not decrease ever in the past. We try
several numbers of the ancestors $M$(0), with sizes $P_i$($0$),
where we assign new random growth rates for the new cities, which
are not changed later, as well as the growth rates for ancestors are
kept the same through the time evolution. It is obvious that
$H$=1=$G$ gives the gradual evolution for the cities, where we have
regular fragmentation with $H$ (and, with some $S$) at each time
step $t$. This case is kept out of the present scope, because we
consider it as (historically) unrealistic. We know that newly
established cities usually continue to speak the language of the
home city, i.e., the official language of their (country) state, at
the present time. Yet, in history, about 1,000 to 1,500 ago, the
situation was different; then there were big states (empires; to
name the main ones: Roman (Byzantine), German, Russian, Ottoman,
Chinese, Japanese), and many citizens of these states were speaking
different languages than the official one of the state, which were
being used mostly in administrative issues, or in written
literature, etc. And these empires were rather interested in
economical and military issues of their members than the cultural
ones. So, many cities could select a language different than the
official language of the state (which would be similar to the home
city that they are fragmented from or a different one). Or, when a
mass immigration takes place (it is known that mass immigrations
were taking place quite often in Europe, within the Middle Age,
because of wars, widespread illnesses, etc.), the newly established
cities were generating new languages in time, instead of continuing
to speak the home language or selecting the language of one of the
neighbor cities. And about some 1,500 to 2,500 years ago (in fact,
there is not enough records about this era; yet, it might be guessed
and the available data may be utilized) the new cities were usually
creating new languages, since (almost) each city was a state (the
generation of a new city was almost the same as the emergence of a
new state, or vice versa). After considering all of these (and other
possible) cases, we take the probability $h_f$ for a new city to
form (create) a new language as $H$/50, where $H$ is the
fragmentation rate for the cities; i.e., $0< h_f \ll 1$. Please note
that the ratio of the number of the living cities (about a million)
to the number of living languages (about 7,000) is about 100. And we
guess that this ratio was much smaller than the present value (about
100) at the beginning (at $t$=0). So, our assumption about $h_f
(\approx H/50)$ may be considered as reasonable. Following similar
considerations, the probability for a newly generated city to select
one of the current languages ($h_s$) is taken as about 0.0001; so,
the probability for a newly generated city to continue to speak her
old language ($h_o$) may simply be taken as $h_o= 1- h_f  - h_s = 1
- H/50 - 0.0001$. It is obvious that $h_o \approx 1$, since $H$ is
small, as we will consider in Sect. 3. More precisely $h_f \approx H
/( M_p / m_p)$, where $M_p$ and $m_p$ is the number of the cities
and that of the languages at the present time, respectively.

    It may be worthwhile to note that the size distribution for the
present languages that resulted from our many runs (not shown) are
found not to be very sensitive to $h_s$, which might be due to the
following reasons: 1) Different cases (which are discussed at the
beginning of the previous paragraph) were important in different
historical time periods (ages, eras, etc.). And in the long run
(i.e., through the history) the effect of changing the languages by
the cities became minor, due to time averaging in the long term,
which may also be related to the following reason: 2) When the ratio
of the number of the present cities to the number of living
languages is considered, the following question may be asked: Why is
the number of the living languages very small with respect to the
number of the living cities? Our answer (our guess) is that many
languages are formed (created) in history (in terms of
fragmentation, etc.) and many of them became useless in time. So,
many languages are formed in time (in history) and many cities
changed their language. And at the end we have about 7,000 living
languages now, which are spoken in about a million living cities. 3)
It is known (also, due to our relevant prediction, which will be
mentioned in Sect. 3) that the world population and the number of
the cities increase exponentially with time. So, the situations in
the past became less important at the present time and the recent
(and the present) ones became dominant, since the number of cities
and the number of the citizens are huge (with respect to history)
now. We think that not the value but the order of magnitude is
important for $h_s$, and we take $h_s$=0.0001. Please note that, in
any case, if a city fragments by $S$, then her language fragments by
a lesser (or occasionally equal) ratio, since the number of speakers
of the home language is greater than (or occasionally equal to) the
population of the new city. Secondly, the fragmentation ratio for
the language may be considered as random, since the splitting city
and her population are selected randomly. And, if all of the cities
(where the given language is spoken) become extinct (due to
fragmentation and probable punctuation), then the given language
becomes extinct.

The time evolution of (the total number of human living in all of
the current cities or speaking all of the current languages, i.e.)
the world population is

$$W(t) = \sum^{M(t)}_{I=1} P_I(t)  =  \sum^{m(t)}_{i=1} p_i(t)  ,
\eqno(2)$$ where a city may actually also be a village or a single
house, except if we demand a minimum population.

If one changes $M(0)$ to $M'(0)$ without changing the origin and the
unit for time scale, then $M(t)$ must be changed to $M'(t) =
M(t)(M'(0)/M(0))$. (Similarly for languages in terms of $m(0),
m'(0), m(t)$, and $m'(t)$.) Secondly, if time steps of the evolution
are long enough (and the total number of time steps is small), we
may have more fragmentation per unit time. If on the other hand,
time steps are rather short (for big number of total time steps),
intermittency may become crucial. Moreover, there may be various
other reasons for waiting periods of time in fragmentation, as well
as in population growth. (That is why we assumed an intermittency
factor $H$ for splitting, the maximum of which is taken the same for
all of the cities.) Thirdly, we know that not all of the ancestor
and old cities (or languages) survive, and many of the old cities
are archeological sites now and we have many ancient (and older)
languages which are not spoken any more. The relevant number of the
people might have immigrated and changed the city and the language,
they might have changed their language after being colonized within
the same cities and they might have become totally extinct after a
lost war, or after a widespread and severe illness, etc. We keep the
(historical) reasons out of the present scope and consider only the
results; and once the reasons are kept away, the results of them may
naturally be taken as random. In summary: for cities, if a random
number (which is defined for each $t$) is smaller than $H$, for
$H<1$, then the city splits (by $S$); and, if a random number is
larger than some $G$ near 1, then the city becomes extinct (and, if
all the relevant cities become extinct then their language becomes
extinct), while the city grows at each time step by $R_I \leq  R$ at
most. Yet, if $H$ is in the order of one part per thousand and if
the number of the time steps is about (few) thousand, it might not
be important whether the growing and fragmentation occur together
with a time step or separately in different time steps. And, if for
$H \ll 1$ this random number is larger than some $G \approx 1$, the
city becomes extinct, where it does not matter whether the cities
become extinct before or after growing.

Please note that what we mean here by extinction of a language is
the extinction of all of the speakers of it. Any (old) language
which is not spoken at the present time may be viewed as a change of
a language by its old speakers, which may be considered as
fragmentation, where splitting may occur in terms of many parts and
rapidly. Namely, a city may fragment and a new city may be
established at the same time. This new city may also form a new
language, which may gain importance with time (as the city gains
importance in terms of commercial relations, political power,
technological developments, etc); this means that many new and old
cities, especially the related ones (sisters, cousins, etc.) may
change their languages. Since many cities take place within the
present event, the situation may be viewed as fragmentation into
many parts; and, since the fragmentation involves many parts, the
relevant language may rapidly become extinct. As a result, a
language becomes extinct but the speakers survive. (This situation
is different from the extinction of a biological taxon in terms of
punctuation, \cite{[5]} where a taxon becomes extinct with the
extinction of all of the members, i.e., the lower taxa and the
species.)

In summary, the cities fragment one by one (if any) at each time
step, where the number of the fragmented cities (at each time step)
may (naturally) be greater than unity. Due to randomness several
cities may fragment into many new ones, within a given time domain
(era, century, etc.). And, many of the new cities may continue to
speak the language of the mother city (cities), a smaller number of
them may select new language(s), where the big languages will be
favored. Thus, a group of new cities may select the same (one of the
big) language(s) within the given time domain; and if this avalanche
continues, the given old language may become extinct, i.e., it may
be changed into (replaced by) another one. This is how the
multiplicative noise and fragmentation may work in shaping the
(current or present) size distribution for the languages, in terms
of groups of the cities. Please note that we consider the evolution
of the languages as a result of the evolution of the cities; and we
assign an ancestor language to each ancestor city, and follow each
city for the (evolution of) languages in terms of growth,
fragmentation and extinction.

\subsection{Lifetimes for cities or languages}

For lifetime, we simply subtract the number of the time step at
which a city or a language is generated, from that one at which the
given agent became extinct, where the agent may become extinct since
its size becomes less then unity in terms of fragmentation or since
the city is randomly eliminated; and if all the cities which were
speaking a given language are randomly eliminated, then we consider
the given language(s) as being randomly eliminated. Please note that
random elimination of languages, which is independent of the cities
amounts to changing the language and within the present approach,
random elimination of languages is considered with the meaning
mentioned in this line. In any case, we have two different
definitions for lifetime; one is for the extinct agents and the
other is for the living agents (age). We may define  $\mu(\tau , t)$
as the probability (density) function, where $t$ stands for the
number of time steps, and $\tau$  stands for ages, and  $\mu_C$
counts the cities (or $\mu_L$ counts languages) with age  $\tau$ at
$t$ (with $0 < \tau \leq t$). It is obvious that the integral of
$\mu_C$ $(\mu_L$) over $\tau$ (with $0< \tau \leq t$) gives the
number of living cities (languages) at $t$, minus those surviving
from beginning, which is equal to the number of total living agents
($M(t=2000), \; m(t=2000)$) minus the number of living ancestors.
The latter goes zero with $t \rightarrow \infty$, due to
fragmentation and (probable) punctuation.

Please note that the introduced parameters (with the symbols in
capital letters for the cities and those in lower case for the
languages) have units involving time, and our time unit is
arbitrary. After some period of evolution in time we (reaching the
present) stop the computation and calculate the probability
distribution (density) function (PDF) for size, and for some other
functions such as extinction frequency, lifetime, etc., (for cities
or languages).  The number of interaction tours may be chosen as
arbitrary (without following historical time, since we do not have
historical data to match with), with different time units; and the
parameters (with units) may be refined accordingly. Yet, in most
cases relative values (with respect to other cases; population
growth rate in different runs, for example) and ratios of the
parameters (the ratio of $S$ to $H$, for example) are important.

\subsection{Family trees for the cities or the languages}

We have $M$($0$) ancestor cities and $m$($0$) ancestor languages,
and they evolve in time, in terms of multiplicative noise and
fragmentation; and, as they fragment new agents emerge, when no
(new) city can be created since no citizen can be created), yet a
(new) language can be created arbitrarily. So, it would be nice to
know how many living languages are the grand children (offspring) of
the ancestors and how many of them are created on the way. It is
obvious that all of the living cities are the offspring of the
ancestor cities. It would also be nice to know from which generation
(level) a given city or a language is, and how these numbers are
distributed over the cities or the languages, etc. So, we need to
know the family trees for the cities and the languages, and we
construct them in the following way: We assume that all of the
initial cities and the initial languages form different families;
i.e., we have $F$($0$) many city families and $f$($0$) many language
families at $t$=0. We label each city by these indices; i.e., the
city family index and the language family index, which may not be
the same. Please note that the family indices increase one by one.
As time goes on, we add a small real number ($\Delta$, which is much
less than the reciprocal of the total number of the time steps) to
the family index of both fragmented cities, when both cities
continue to speak the old language. So, we have the family numbers
close to the original index values (yet little different from them)
at the end. If any city forms a new language, which is fragmented
from the old one, then we add $\Delta$  to her language family
number. If the new city changes her language and selects one of the
current languages, then we replace her language family number with
that of the selected language. And at $t$=2,000, we truncate the
family numbers and count how many cities have the given index (1, 2,
3, ..., etc.) from her city family number and language family
number. Furthermore, by dividing the residuals by $\Delta$, we
compute the generation number(s) (level). In this manner, we are
able to compute the number of the members of each family, as well as
their sizes at the present time, etc.

As we considered at the beginning of the present section, no city
may be created out of nothing yet a language may be created out of
nothing, since many words or a grammatical rules can be created
arbitrarily, i.e., a single person or a small group of people may
agree on an arbitrary language, which is not fragmented from any
other language. But, when they establish a city, this means that the
old city (that they were the inhabitants of) is fragmented. It is
obvious that the unification (merging) of the cities and the
languages are kept out of the present scope.

\bigskip
\section{Applications and results}

It is well known that the cities and the languages are man made
systems; and a city is a physical quantity whereas a language is not
a physical quantity. Yet, they are connected and we find many
similarities within the relevant results presented here.

Empirical criteria for our results are: i) The number of the living
cities (towns, villages, etc.) and that of the living languages may
be different; but, total size for the present time must be the same
for both cases, where the mentioned size is the world population
(Eqn. (2)). ii) World population increases exponentially (super
exponentially, for recent times) with time.\cite{[6]} iii) At
present, the biggest language (Mandarin Chinese) is used by about
1.025 billion people and world population (prediction made by United
Nations) is 6.5 billion in 2005, (and to be about 10 billion in
2050) \cite{[6]}; so the ratio of the size for the biggest language
to (the total size, i.e.,) world population must be (about) 1:6.5.
iv) The size distribution for the present time must be a power law
--1 for the cities (Pareto-Zipf law), and it may be a slightly
asymmetric log-normal distribution for the languages. We first
consider cities (Sect. 3.1), later we study languages (Sect. 3.2).

\subsection{Cities}
{\it Initialization:} In our initial world (at our $t$=0) we have
$M$($0$) (=1,000) many cities, each of which is inhabited by
randomly chosen $P_I(0) \; (\leq 1,000$) people; thus, the initial
world population ($W(0)$) is about 500,000, since the average of
homogeneous random numbers between zero and unity is 0.5. Thus we
assume power law zero for the initial distribution of languages over
size.

It is obvious that we may not set our time origin correctly, because
no real data for the initial time is available to match with. Yet,
we may assume a homogeneous distribution of size over initial
cities. In fact, we tried many smooth (Gauss, exponential, etc.)
initial distributions (not shown); and all of them undergo similar
time evolutions within 2,000 time steps, under the present processes
of random multiplication for growth, and (random) fragmentation for
spread and origination and extinction, where we utilized various
combinations of the relevant parameters, including $H$ and $G$. We
tried also a delta distribution, which is equivalent to assuming a
single ancestor, for the initial case; it also evolved into a power
law about --1 (with a different set of parameters, not shown) in
time. In all of them, we observe that the city distribution at
present (Pareto-Zipf law) is independent of initial (probable)
distributions, disregarding some extraordinary ones.

Please note that our initial conditions, which are introduced within
the previous two paragraphs, may be considered as corresponding to
some 10,000 years ago. So the unit for our time steps may be taken
as (about) 5 years, since we consider 2000 time steps for the
evolution of cities in the following sections.

\textit{Evolution without elimination} ($G=1$): As $t$ increases,
the cities start to be organized; and within about 200 time steps,
we have a picture of the current world which is similar to the
present world, where the distribution of city populations is
considerably far from random. With time, the number of cities
($M(t)$) and the world population ($W(t)$) increases exponentially,
as shown in Figure 1, where the exponent for cities (which decreases
with $t$) is 0.0014, and that for world population (which increases
with $t$; dashed line) is 0.0024, (and for $t$=2000, in all) where
the parameters are: $R$=0.0078 and $H$=0.004, with $S$=0.49999
(others are same as before). Please note that at $t$=2000 (present
time, the year 2000) we have about half million cities and the world
population is about 17 billion (Fig. 1).

Figure 2 is the time evolution of the size distribution of the
cities (PDF), all of which split and grow by the same parameters,
i.e., $S$, $R$, $H$ (all same as declared in the previous
paragraph), and $G$=1 (thus, we do not have abrupt (punctuated)
elimination of cities here). The lower plot is historical ($t$=320,
open squares) and upper one (solid squares) is for the present time
($t$=2000). Please note that in Fig.2 the arrow has the slope --1,
which indicates the Pareto-Zipf law for the cities. Furthermore, we
observe that as initial cities spread in number by fragmentation,
the initial random distribution turns out to be Gauss for
intermediate times (as the parabolic fit indicates, for $t$=320 for
example) which becomes a power law --1 (at the tail, i.e., for big
sizes) for the present time.

\textit{Elimination} (punctuation, $G<1$) plays a role, which is
opposite to that of fragmentation ($H$) and growth ($R$) in
evolution; here $H$ and $R$ develop the evolution forward, and $G$
backward. So the present competition turns out to be the one between
$H$ and $R$, and $G$, where two criteria are crucial: For a given
number of time steps, $R$, and $M$($0$), etc., there is a critical
value for $G$; where, for $G_{critical} < G \approx 1$ cities
survive, and for smaller values of $G$ (i.e., if $G \approx
G_{critical}$) cities may become extinct totally. (For similar cases
in the competition between species, one may see  \cite{[5]}, and
references therein.) Secondly, sum of $H$ and $G$ is a decisive
parameter for the evolution: If for a given $G$ (with $G_{critical}<
G$), $H$+$G$=1, then the number of cities does not increase and does
not decrease, but oscillates about $M(0)$, since (almost) the same
amount of cities emerges (by $H$) and becomes extinct (by $G$) at
each time step, and we have intermediate elimination. On the other
hand, if $H+G<1$, then the cities decrease in number with time and
we have high (strong, heavy) elimination. Only for $1< H + G$, (with
$G \neq 1$) we have low (weak, light) elimination of cities, where
the number increases (yet, slowly with respect to the case for
$G$=1). It is obvious that elimination slows down the evolution. It
amounts to decreasing the unit for a time step, and increasing the
number of the time steps from the initial time ($t$=0) up to the
present, and vice versa. So we need more time to arrive at a given
configuration of the cities; i.e., total number, size, total size
PDF, etc. In summary, only light punctuation of the cities may be
historically real, and it does not affect the evolution and size
distribution of the cities, as we observed in many runs (not shown),
where we increase the fragmentation ($H$) and population growth rate
($R$) to compensate the negative effect of punctuation on the number
of cities and world population, respectively. Yet, the ancestor
cities i.e., those at age of $t$ at any time $t$, decay more quickly
in time as (punctuation increases) $G$ decreases (since the
generated cities may be substituted by newly generated ones after
elimination; but the ancestor ones can not be re-built.)

In summary, only light punctuation of the cities (together with all
of the citizens) may be historically real, as many (deplorable)
examples occurred during many wars, and it does not affect the
evolution and size distribution of the cities, as we observed in
many runs (not shown), where we increase the fragmentation ($H$) and
population growth rate ($R$) to compensate the negative effect of
punctuation on the number of cities and the present world
population, respectively. 

\textit{Lifetimes}: Time distribution of cities (lifetime for
extinct cities, and ages of the livings ones) are decreasing
exponentials (disregarding the cases for small number of ancestors
and high punctuation) as shown in Figure 3. Simple probability
(density) functions are also exponential (not shown), which means
that cities occupy the time distribution plots in exponential order;
more cities for small $t$, and fewer cities for big $t$, for a given
number of time steps in all. We predicted that the (negative)
exponent of the simple probability (density) function of the living
cities at $t$=2,000 is about 0.0015 per time step as shown within
the inset of Fig. 5, where the number of living ancestor cities
($M$($0$)=1,000) could be neglected within about a million of the
cities at present if the ancestors have small sizes. Within the
probability (density) function  $\mu_C(\tau,t)$, $t$ stands for the
number of time steps, and $\tau$ stands for ages of the cities at
$t$, and $\mu_C$ counts the cities with age at $t$ (with $0< \tau \leq
t)$. Please note that we have only 30 living cities, which are
exactly 1999 years old, one city is 1995 years old, one city is 1990
and one is 1886 years old, etc.

In the inset of Fig. 3 (and in many similar ones, not shown) we
observe that the exponent ($\alpha$, say) of $\mu_C(\tau,t)$
decreases with $t$ (where the number of cities increases with $t$).
Assuming that the exponent ($\alpha =0.0015$) will remain constant
for $t \rightarrow \infty$, we may make a prediction about $M(t
\rightarrow \infty)$ as follows: A simple integration of
$\mu_C(\tau,t) = \mu_C(1,t)\exp(-\alpha\tau)$ over $\tau$ (with $0
<\tau \leq t$) can be performed for $M(t)$, since living ancestors
decay (in terms of fragmentation and random elimination) with time
($M(t)=(\mu_C(1,t)/\alpha)\exp(-\alpha t))$. The crucial parameter
for the result is $\mu_C(1,t)$, i.e. number of newly established
cities at $\tau=1$, for each $t$. It is obvious that $\mu_C(1,t)$
cannot be constant or a decreasing function with $t$, because we
have (light) elimination with $G <1$ in reality, and the number of
eliminated cities increases with $t$ (cities become old and ruined
in time). $\mu_C(1,t)$ may be an increasing function in time, due to
non-zero fragmentation ($H$). If $\mu_C(1,t)$ increases linearly
with time we get saturation in $M(t \rightarrow\infty)$, which is no
contradiction, since we have elimination (in reality). If $\mu_C(1,t)$
increases exponentially with the exponent $\beta$, we have
saturation in $M(t\rightarrow\infty)$ for $\beta=\alpha$; and we
have $M(t \rightarrow \infty)=0$ for $\beta<\alpha$. Since we know
that at present $M(t)$ increases with time, we may guess that
$\alpha<\beta$; i.e., we establish new cities, which increase either
linearly or exponentially (with the exponent bigger than 0.0015 for
cities) in time. One may guess that $M(t \rightarrow\infty$) will
saturate in reality, since human population and surface area of
earth is limited.

On the other hand, it is known that cities were states in the past;
yet, it is not certain whether one language was spoken per city
(state), or vice versa. For the present time, we have about 7,000
languages for a world population which about 10 billion (as
mentioned within the first paragraph of sect. 3.1), and we have
about a million cities. So, one may say that a language is spoken
within 100 (to 500, or so) cities on the average, at present. Yet,
we have many metropolitan cities in the present world, where only
one language is spoken by big majorities. And, we have many cities
per country, where the number of languages is smaller than the
number of cities, etc. Certainly the ratio of the number of
languages to that of cities is less than unity now.

\subsection{Languages}
It is not hard to make the guess that there were many simple
languages (composed of some fewer and simple words and rules), which
were spoken by numerous small human groups (families, tribes, etc.)
at the very beginning. And, as people came together in towns, these
primary languages might have united. Yet, we predict that the
initial world is not (much) relevant for the present size
configuration of the languages (as well as in the case for cities;
see, Sect. 3.1). Moreover, we may obtain similar target
configurations for different evolution parameters (not shown).

Within the present approach, ancestor cities and ancestor languages
are associated randomly, which is what might have happened indeed in
the past. Namely, the languages with their words, grammatical rules,
etc., might have been formed randomly (\cite{[4]}, and references
therein); the societies grew and fragmented randomly (see the last
lines in the fifth paragraph of Sect. 2); new cities randomly formed
new languages or changed their language and selected a new one
randomly. We predict that $i$ (roughly) decreases as $I$ increases
for small $I$ (not shown). We predict also the distribution of the
present languages over the present cities, where we have power law
of minus unity (not shown). It may be worthwhile to remark that
younger cities prefer younger languages; which means also that new
cities (or the new countries which are composed of the new cities)
emerge mostly with new languages. Secondly, as $t$ increases $I$ and
$i$ increase, and the plot of $I$ versus $i$ extends upward and
moves rightward; since, the number of the current languages $m(t)$,
and the number of cities which speak a given language, increase as a
result of the fragmentation of the cities. Furthermore, we computed
the number (abundance) of the speakers for the present languages
($p_i$($t$), in Eq. (2)) (not shown), where we have a few thousand
($m(t=2000)=7587$) living languages, and the exponent (the slope of
the lowest plot) in Fig.1 for the languages is 0.0008. Within this
distribution of the present languages over the speakers, we predict
power law minus unity (not shown). It may be worthwhile to remark
that older languages have more speakers; and in reality (Mandarin)
Chinese, Indian, etc., are big and old languages. For example, we
have about one billion people using the language number 1, which is
one of the oldest languages of the world; and less people speaking
the language number 2, etc.

In Figure 4, we display the PDF for size distributions of the
languages at $t$=500 (historical, open squares) and $t$=2000
(present, squares), where the number of ancestor languages $m(0)$ is
200 (we plotted several similar curves for $m(0)$=1,000 i.e., for
the case where only one ancestor language is spoken in each ancestor
city and obtained similar results; not shown). Splitting rate and
splitting ratio for languages are not defined here, since languages
split as a result of splitting of the cities; and the splitting
ratio of the splitting language comes out as the ratio of the
population of the new city (which creates a new language) to the
total population of the cities which speak the fragmented language.
Please note that Fig. 4 indicates the time evolution of the language
size distribution without any elimination (punctuation), where the
initial distribution (for $t$=0, not shown) is a power law with
exponent zero (yet, this is not much relevant, since we have few
data). In the plot for the present time (squares for $t$=2000) we
have a slightly asymmetric Gaussian for big sizes as the parabolic
fit (dashed line) indicates; and we have an enhancement for the
small languages, which is mainly due to large number of small cities
as also indicated in Fig. 2. It is worthwhile to note that the
empirical size distribution of the present languages \cite{[4]}
displays an hallmark for a power law (roughly minus two) for big
size (for populations bigger than 1,000); the plot decreases almost
linearly (up to some fluctuation) at the big size end, where the
slope is roughly minus two. On the other hand, the small size end
(for the number of speakers less than 1,000) may be random, and it
may involve big fluctuations from time to time, since it may not be
easy to discover (in South Asia, Central Africa, etc.) and count
small languages with the mentioned size, and their number may
(abruptly) change from time to time.

We guess that (random) elimination of languages (with all of its
speakers) is not realistic, and it is not recorded in the history
for the recent times. On the other hand, changing (replacement) of a
language by another one may be realistic; yet, this situation is not
much relevant for the present size distribution. So, in case of a
(hypothetical) random elimination (light), the fragmentation rate
may accordingly be increased to obtain the empirical data for the
present time; and since the names for the languages are irrelevant
for the present formalism, we may totally ignore the punctuation in
the evolution of languages, and consider only the change of the
language(s).

Figure 5 shows the lifetimes for old and present languages (at
$t=180$, and $t=2000$, respectively) where we have exponential
(decreasing) distribution as in case of cities (Fig. 3). The simple
probability distribution functions (not shown) are also (decreasing)
exponentials, as the one for $t=2000$ (inset of Fig. 3) indicates,
where we have the (negative) exponent (=0.0007) as independent of
time. So, following a similar analysis (performed for cities, at the
end of Sect. 3.1) we may predict for the languages that we may have
saturation in number as $t \rightarrow \infty$.

Thus far, we considered that the languages and the cities (as
products of the same humans) are very similar to each other in many
respects. Yet, languages and cities are no longer the same; and for
some intermediate times the cities obey the 1/size law, while the
languages obey the log-normal distribution (or power law minus two
for big size). Cities and languages have (decreasing) exponential
time distributions (lifetime for extinct cities and age for the
livings ones).

In short, the present regularities within the relevant quantities
about the cities and the languages may be the direct result of the
randomness involved, where the two opposite processes
(multiplicative noise and fragmentation) may be organizing the
declared randomness into the observable regularities. (See also
\cite{[5]} for similar cases, where the randomness is evolved into
several regularities (symmetries) in terms of simple averaging
process.) It is worthwhile to underline that emergence of
regularities from randomness may be fundamental and universal.

\subsection{Families of the cities or the languages}

Obviously we do not know how the city (language) families are
distributed over the cities (languages) initially, i.e., at $t$=0,
since we do not have any historical record about the issue. Yet, we
predicted that initial conditions for the cities (languages) are
almost irrelevant for our results. And we considered several initial
conditions for the city families and the language families, which
are (keeping all other parameters as before): 1) Each ancestor city
(language) is considered as a root for the evolution of the city
(language) families; i.e., $F$($0$)=$M$($0$)=1000 and
$f$($0$)=$m$($0$)=200. 2) Each ancestor city (language) is
considered as a root for the evolution of the city families
($F$($0$)=$M$($0$)=1000); and we have $m$($0$)=200 with
$f$($0$)=120, where we assume that the number of the language
families at present \cite{[4],[8]} and in the past ($t$=0, for
example; i.e., about 10,000 years ago in real time) is the same.
Here we do not think about any historical reason for the extinction
of big families (with many cities and languages, and with big
populations); yet some limited number of small families might have
emerged on the way.

In 1) and 2) here we have the city families and the language
families as disconnected, which might be a false assumption. For
people speaking certain languages are the citizens of certain cities
(countries), and related languages are spoken in related cities
(countries) at the present time, which was similar in history. For
that reason, we coupled the initial city families and the language
families as: 3) We considered $M$($0$)=1000=$m$($0$) (after lowering
the fragmentation rates for the language suitably, on the aim of
obtaining the present number of the languages and the relevant size
distribution); and we assumed that the initial city families and the
language families are the same; i.e., $F$($0$)=120=$f$($0$). 4)
Similarly we considered $M$($0$)=$m$($0$)=1000=$F$($0$)=$f$($0$). We
utilized $\Delta \ll 1/2,000$ (since we have 2,000 time steps), more
precisely $\Delta=0.0001$, (with no punctuation, $G=1$) in all
(i.e., 1 to 4)).

We obtained similar plots under all of the present assumptions about
the initial conditions for the families (not shown); and the Figures
6, 7 are for the language families under 3); here the families are
placed on the horizontal axes with the rank in decreasing order
(i.e., the biggest family (in number in Fig. 6, and in size in Fig.
7) has the rank 1, the next biggest family has the rank 2, etc. It
is clear that the distribution of the present language families over
the number of the present languages displays decreasing exponential
behavior in Fig. 6 (except at the big and small number ends). It is
known that the data for the present language families \cite{[4]} may
be interpreted as involving a power law about --2 \cite{[8]}. Yet,
another data for the present language families \cite{[9]} seems to
involve no power law at all (please see also \cite{[8]}, for the
relevant comparison). And our related prediction may be considered
as similar to the one in \cite{[9]}, as the log-log plot displays in
the inset of Fig. 6.

Fig. 7 is for the size distribution of the present language
families, which may be considered as being in good agreement with
the ones in \cite{[4],[8],[9]}, as the corresponding log-log plot
designates in the inset of Fig. 7.

\bigskip
\section{Discussion and conclusion}

Starting with random initial conditions and utilizing many
parameters in two random processes for the evolution, we obtained
several regularities (for size and time distributions, etc.) under
the process of random multiplicative noise for growth and random
fragmentation for generation and extinction of languages or cities.
We find that results are (almost) independent of initial conditions,
disregarding some extra ordinary ones. If we have $G \neq 1 \; (G
\approx 1$), we need longer time to mimic a target configuration,
where new generated cities or languages may be inserted, in terms of
fragmentation, provided $G_{critical} \leq G$, and $1\leq G + H$.
Furthermore, many cities or languages become extinct in their youth,
while less of them become extinct as they become old. In other
words, languages or cities become extinct either with short lifetime
(soon after their generation), or they hardly become extinct later
and live long (which may be considered as a kind of natural
selection).

Considering our results (which are displayed in Sect. 3, and within
the relevant figures), we may state that the present processes
(namely, fragmentation and multiplicative noise) with the relevant
parameters ($H$ for the fragmentation, and $R$ for the
multiplicative noise process; for the cities, for example) give
reasonably good results, in terms of which we may understand and
explain the present situation for the cities or for the languages
and their time evolutions. Yet, our assumptions about the initial
world and several of our explanations (which may be considered as
probable) about the reasons during evolution may or may not be true,
since we do not have much information about these periods of time,
due to limited data about historical records. In fact, (as we
mentioned in the first paragraph of Sect. 4 that) the initial
conditions are not much relevant for the results, since the origin
of our time and the relevant number of time steps (and other
parameters) may be selected accordingly. Now, we may claim that
(whatever the historical reasons were in reality) the cities and the
languages grew in size and they fragmented, and these realities are
covered within the present approach as the essential issue(s).

It is known that logistic maps (Eq. (1)) are dynamic, and under some
circumstances they may become chaotic. We predict that time
distribution for the cities or the languages are (decreasing)
exponential, and log-normal size distribution for languages at
present may turn into a power law in future. The power law --1 for
size distribution of the present cities, may be the result of
(decreasing) exponential distribution of their probability over
size. In summary, cities and languages are not the same and for some
intermediate times the cities obey the log-normal distribution;
which may be checked within archeological data (as a subject of a
potential field of science; we suggest its name as physical history) for 
ancient cities (towns) and their population.

The author would like to thank to D. Stauffer and to an anonymous
referee for many discussions and corrections, and for many other
reasons.


\vskip 2cm
\begin{figure}
\begin{center}
\includegraphics[scale=0.9]{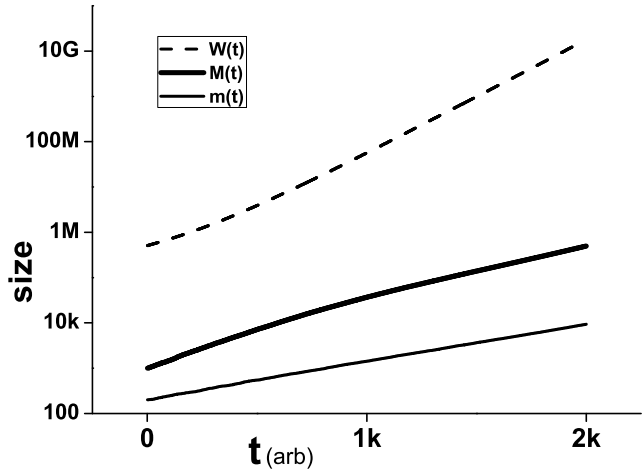}
\end{center}
\caption{Time evolution of world population $W(t)$ (dashed), of the
number of the cities $M(t)$ (thick solid) and of languages $m(t)$
(thin solid) (Eqn. (2)). Exponent of $W(t), \; M(t)$, and $m(t)$ is
0.0024, 0.0014, and 0.0008, respectively. Please note that the
vertical axis is logarithmic.}
\end{figure}

\begin{figure}
\begin{center}
\includegraphics[scale=0.6]{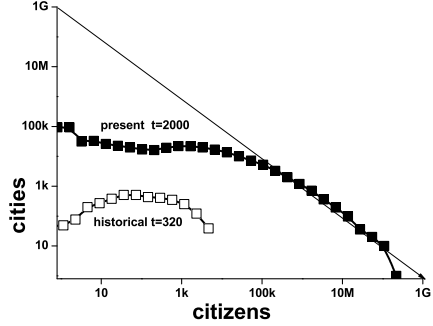}
\end{center}
\caption{Historical (open squares, for $t$=320) and present
(squares, for $t$=2000) size distribution of the cities. (Parameters
are given in sect. 3.1.) Please note that the city with maximum
population is within the last bin and we have about 10 biggest
cities at $t$=320, since the horizontal axis is adjusted to involve
the present biggest cities with about hundred million population. We
have 3657 historical cities, and 497,386 living cities.}
\end{figure}

\begin{figure}
\begin{center}
\includegraphics[scale=0.6]{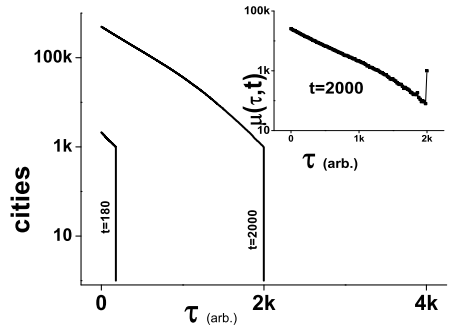}
\end{center}
\caption{Histogram of city ages $\tau$ in the past ($t=180$) and at
present ($t=2,000$). (Parameters are given in sect. 3.1.) Please
note that the number of generated (new) cities decreases
exponentially with age, in all cases, and as $t$ increases (number
of cities increases and) the exponent (of $\mu_C$) decreases. Many of
the ancestor cities live at any $t$, and few of them becomes extinct
in terms of fragmentation (for the present time, about hundred). The
dot within the plot of the inset represents the living ancestors
($\approx 1,000$).}
\end{figure}

\begin{figure}
\begin{center}
\includegraphics[scale=0.6]{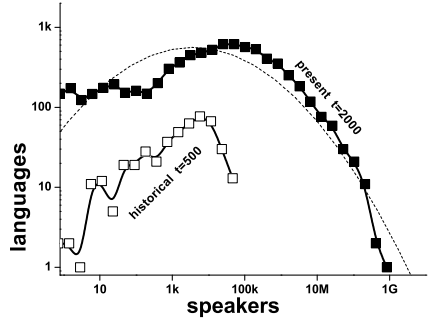}
\end{center}
\caption{Size distributions of languages: Open squares are for the
past ($t$=500) (where we have 456 languages, with $m(0)=200$) and
solid squares are for the present ($t$=2,000) (where we have 7,587
languages, with $m(0)=200$). (See, Sect. 3.2 for the relevant
parameters.) For the empirical data, please see [4-6]. Similar
results, also for city sizes, were seen for the future at $t =
3,000$.}
\end{figure}

\begin{figure}
\begin{center}
\includegraphics[scale=0.6]{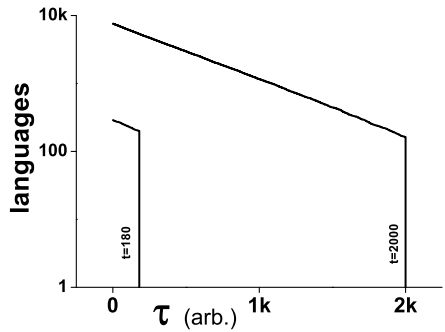}
\end{center}
\caption{Time distribution for (extinct and living) languages. (See
Sect. 3.2 for the parameters.) Please note that the (negative)
exponent (of $\mu_L$) is time independent.}
\end{figure}

\begin{figure}
\begin{center}
\includegraphics[scale=0.6]{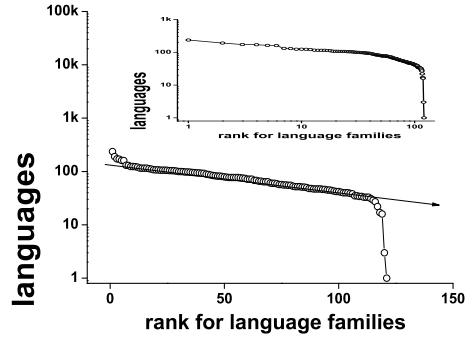}
\end{center}
\caption{The language families (in rank order along the horizontal
axis, which is linear) versus their number of languages (along the
vertical axis, which is logarithmic) at the present time, where we
have 120 families. Decreasing exponential trend is clear, except for
the small families. The inset is the same plot, in log-log scale.}
\end{figure}

\begin{figure}
\begin{center}
\includegraphics[scale=0.6]{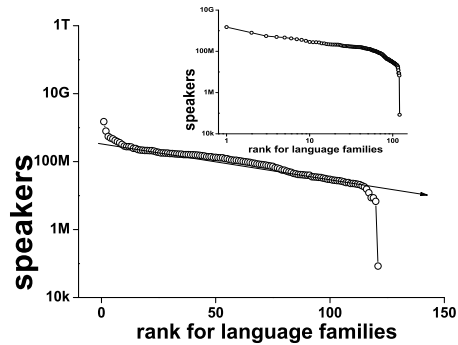}
\end{center}
\caption{The language families (in total 120, in rank order along
the horizontal axis, which is linear) versus their number of
speakers (along the vertical axis, which is logarithmic) at the
present time. Decreasing exponential trend is clear, except for the
small families with small languages as the members. The inset is the
same plot, in log-log scale. }
\end{figure}
\end{document}